# Generic Automatic Proof Tools[1]

*Lawrence C. Paulson*
Computer Laboratory, University of Cambridge

May 1996

---

# Contents

# 1 Introduction

This article explores a synthesis between two distinct traditions in automated reasoning: resolution and interaction. In particular it discusses Isabelle, an interactive theorem prover based upon a form of resolution. It aims to demonstrate the value of proof tools that, compared with traditional resolution systems, seem absurdly limited. Isabelle's classical reasoner searches for proofs using a tableau approach. The reasoner is *generic*: it accepts rules proved in applied theories, involving defined connectives. The reasoner works in a variety of domains without reducing them to first-order logic.

Resolution systems such as Otter [13], SETHEO [11] and PTTP [34] represent automatic theorem proving at its highest point of refinement. They achieve extremely high inference rates and can run continuously for days without running out of storage. They can crack many of the toughest challenge problems that have been circulated. While they exploit many specialized algorithms, data structures and optimizations, they rely crucially on unification.

Interactive systems let the user direct each step of the proof. They can implement complicated formalisms, chosen for maximum expressiveness, and typically based on the typed $\lambda$-calculus. HOL [7, 8] and PVS [23] are used for verification of hardware and real-time systems, while Coq [4] is used for formalizing mathematics. Large numbers of axioms — say, the description of a CPU design — do not overwhelm them, because finding the proof is the user's job. Partial automation is sometimes provided, but a resolution enthusiast would regret the lack of uniform search procedures based on unification.

One procedure provided by most interactive provers is rewriting. Rewrite rules have many advantages. Unlike programmed inference rules, they are declarative; to replace terms of the form $t - t$ by 0 we supply the rule $x - x = 0$. As rewrite rules are not translated into anything like clause form, we can follow the reasoning. Apart from conditional rewrites such as $x \neq 0 \rightarrow x/x = 1$, rewriting does not involve search. Certain rewrites, like the distributive law $(A \cap B) \cup C = (A \cup C) \cap (B \cup C)$, can increase the effort needed to find a proof, but there is no search space to explode.

Isabelle [27] is an interactive prover based on the typed $\lambda$-calculus. But its primary inference rule is a generalization of Horn clause resolution. Isabelle uses resolution to implement proof checking in a generic (logic-independent) way. It provides no uniform search strategy, but sev-eral tools based on lazy lists. They can express depth-first, best-first and iterative-deepening strategies, and can be combined to yield automatic proof procedures. One such procedure is the classical reasoner. It has many of rewriting's advantages: it applies a user-supplied set of rules in a declarative manner, usually with a small search space.

Automatic provers have great power, while interactive systems can support sophisticated formalisms. PVS and Isabelle illustrate two distinct ways of adding power to interactive systems. PVS adds decision procedures, while Isabelle exploits unification and tableaux. Decision procedures are highly efficient, but specialized; tableau methods provide significant automation over many domains. The two approaches are complementary and can be used in a single system. Decision procedures are fairly well described in the literature [3]; this article attempts to do the same for Isabelle's approach.

**Outline.** After this introduction (§1), the article presents Isabelle: theory, proof checking, and search (§2–5). It then reviews the tableau approach (§6). It describes Isabelle's classical reasoner (§7), with simple examples in first-order logic and set theory. A major example, the Church-Rosser theorem for combinators, is introduced (§8) and its proof in Isabelle described (§9–11). The conclusions (§12) include statistics derived from proof scripts, which indicate that the classical reasoner is used about as heavily as the rewriter.

## 2 Isabelle

Isabelle's original design goal was to provide a combination of interaction and unification. Unification's benefits are familiar. First, it allows a proof search to proceed while certain terms remain unspecified. Second, it supports the extraction of answers from proofs. The second point, initially of interest to artificial intelligence, is also relevant to the "proofs as programs" school of program verification.

During implementation, a third advantage became evident. By adopting Horn clause resolution as a primitive inference, the prover could be made *generic*: it could support proof in any logic whose inference rules could be expressed as Horn clauses. As I was already working with Horn clauses generalized to a higher-order formalism, the approach could handle all conventional quantifier rules and thus cover a rich variety of logics.

Figure 1 displays some of the logics that have been implemented using Isabelle. The automatic proof tools described in this article apply only to the classical natural deduction logics FOL and HOL (higher-order logic) and their descendants such as ZF set theory. Analogous but less developed tools exist for the classical sequent calculus LK. Intuitionistic logics such as IFOL and CTT require different techniques, but unification and search remain the key to automation.

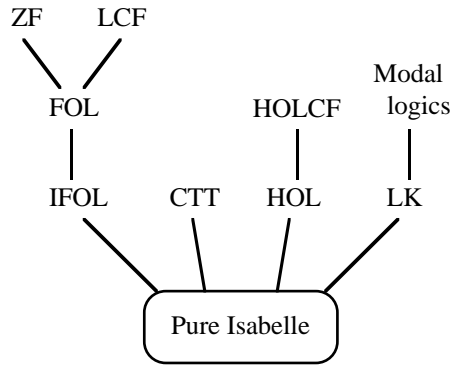

Figure 1: Some Isabelle Logics

Sometimes the term *generic* is applied to any prover whose logic is general-purpose. Specialized formalisms can be embedded by adding suitable axioms, as is frequently done in provers for higher-order logic. But reasoning in the embedded logic is not easy; Pelletier [31, page 209] notes that such embeddings are a good way of generating challenge problems for theorem provers.[1]

Isabelle consists of a *meta-logic* packaged so as to support reasoning in embedded logics, which are called *object-logics*. These include IFOL and HOL, as well as logics embedded in them. The underlying typed $\lambda$-calculus provides excellent support for HOL. But it is beneficial even in first-order logics, such as ZF: new binding operators, such as $\bigcup_{x \in A} B(x)$, can be defined easily.

---

[1] Pelletier attributes the idea to Charles Morgan.

# 3 A Glimpse at the Theory

Let us examine how object-logic rules can be expressed as generalized Horn clauses. Natural deduction typically defines conjunction by these rules:

$$\frac{\phi \quad \psi}{\phi \wedge \psi} \quad \frac{\phi \wedge \psi}{\phi} \quad \frac{\phi \wedge \psi}{\psi}$$

The corresponding Horn clauses are

$$[\![\mathbf{Tr}\phi; \mathbf{Tr}\psi]\!] \Longrightarrow \mathbf{Tr}(\mathbf{And}\phi\psi) \qquad (\wedge I)$$

$$\mathbf{Tr}(\mathbf{And}\phi\psi) \Longrightarrow \mathbf{Tr}\phi \qquad (\wedge E1)$$

$$\mathbf{Tr}(\mathbf{And}\phi\psi) \Longrightarrow \mathbf{Tr}\psi \qquad (\wedge E2)$$

Here **Tr** is a predicate to identify the true formulæ, while **And** is a function representing conjunction. To Isabelle, both symbols are uninterpreted.

In general, the inference rule

$$\frac{\phi_1 \quad \phi_2 \quad \cdots \quad \phi_n}{\psi}$$

is represented by a clause of the form $[\![\phi_1; \phi_2; \ldots; \phi_n]\!] \Longrightarrow \psi$. The brackets $[\![$ and $]\!]$ are optional if $n = 1$; this syntax merely abbreviates

$$\phi_1 \Longrightarrow (\phi_2 \Longrightarrow \cdots (\phi_n \Longrightarrow \psi) \cdots).$$

The connective $\Longrightarrow$ is meta-level implication.

Now consider the natural deduction rules for implication:

$$\frac{\begin{array}{c}[\phi]\\ \psi\end{array}}{\phi \to \psi} \quad \frac{\phi \to \psi \quad \phi}{\psi}$$

The first rule discharges the assumption $\phi$; the corresponding Isabelle clause expresses this by nesting $\Longrightarrow$ to the left. This is no ordinary Horn clause. The other rule is straightforward, and **Imp** represents FOL's implication symbol.

$$(\mathbf{Tr}\phi \Longrightarrow \mathbf{Tr}\psi) \Longrightarrow \mathbf{Tr}(\mathbf{Imp}\phi\psi) \qquad (\to I)$$

$$[\![\mathbf{Tr}(\mathbf{Imp}\phi\psi); \mathbf{Tr}\phi]\!] \Longrightarrow \mathbf{Tr}\psi \qquad (\to E)$$

Finally, consider the natural deduction rules for universal quantification:

$$\frac{\phi}{\forall x\,\phi}\ *\qquad \frac{\forall x\,\phi}{\phi[t/x]}$$

Here $*$ stands for the proviso "$x$ not free in assumptions", while $\phi[t/x]$ is the result of replacing $x$ by $t$ in $\phi$. The proviso ensures that $\phi$ is proved for arbitrary $x$. The corresponding Isabelle clause expresses this using the meta-level universal quantifier, !!. Again, this is no ordinary Horn clause.

$$(!!x.\,\mathbf{Tr}(\phi x)) \Longrightarrow \mathbf{Tr}(\mathbf{All}\phi) \qquad (\forall I)$$
$$\mathbf{Tr}(\mathbf{All}\phi) \Longrightarrow \mathbf{Tr}(\phi t) \qquad (\forall E)$$

Now $\phi$ has become a function variable. The substitution shown in the original rules is handled by function application and $\beta$-reduction in Isabelle's $\lambda$-calculus. Similar techniques handle the existential quantifier and induction rules [24].

Isabelle's meta-logic consists of the connectives !! (for all), $\Longrightarrow$ (implies) and $\equiv$ (equals). Miller's $\lambda$Prolog [15] has a similar logic and clauses, and generalizes resolution in a similar way. There is one crucial difference. $\lambda$Prolog is a logic programming language, with a control strategy based upon Prolog's; theorem provers can be coded in it [5]. Isabelle uses resolution to support proof checking, with no built-in strategy; logic programming effects can be obtained, but virtually all programming is done using ML [29].

## 4  Isabelle as a Proof Checker

Some interactive sessions (using Isabelle/HOL) will illustrate these concepts. To begin, let us prove the tautology $P \rightarrow (Q \rightarrow P)$. We direct Isabelle via the ML top level:

```
goal HOL.thy "P --> (Q --> P)";
  Level 0
  P --> Q --> P
   1. P --> Q --> P
```

The output (in *typewriter italics*) is the initial goal clause. It has the form $G \Longrightarrow G$, where $G$ is our tautology.

In general, Isabelle goal clauses have the form $[\![G_1;\ldots;G_n]\!] \Longrightarrow G$, where $G$ is the ultimate goal to be proved. Proving $G_1, \ldots, G_n$ therefore proves $G$ within the system. Contrast this with classical resolution, which uses **false** instead of $G$ and aims to prove a contradiction.

Resolving the $(\rightarrow I)$ clause with the goal clause yields a new goal clause. The head is as before; the body is the new subgoal:

```
by (resolve_tac [impI] 1);
  Level 1
  P --> Q --> P
   1. P ==> Q --> P
```

The subgoal has the form $\mathbf{Tr}P \Longrightarrow \mathbf{Tr}(\mathbf{Imp}QP)$. where $\mathbf{Tr}P$ represents the assumption $P$. We need to apply $(\rightarrow I)$ a second time. Isabelle's generalized resolution "lifts" the clause over the assumption, copying it to all resulting subgoals:

```
by (resolve_tac [impI] 1);
  Level 2
  P --> Q --> P
   1. [| P; Q |] ==> P
```

The new subgoal asks us to prove $P$ given assumptions $P$ and $Q$. Isabelle can recognize this as true by assumption.

```
by (assume_tac 1);
  Level 3
  P --> Q --> P
  No subgoals!
```

The resulting goal clause is $P \rightarrow (Q \rightarrow P)$, which is the theorem we set out to prove. We could now store it for use in future proofs.

Resolution and proof by assumption are derived inference rules of the meta-logic and follow by the usual properties of !! and $\Longrightarrow$. From now on, let us adopt a more readable syntax for clauses, suppressing the symbol **Tr** and using conventional logical notation.

This trivial example illustrates how resolution can support proof checking. We refer to object-logic rules such as $(\rightarrow I)$ by ML identifiers such as `impI`. It does not matter whether the rules are taken as primitive or are derived from other rules. Most proof checkers provide a separate command for each rule; adding new rules often involves programming.

For a second example, let us prove $(\forall x\, Px) \rightarrow \forall y\, (Qy \rightarrow \forall z\, Pz)$. Note that Isabelle's quantifier notation incorporates a dot, giving quantifications the largest possible scope:

```
goal HOL.thy "(ALL x. P x) --> (ALL y. Q y --> (ALL z. P z))";
  Level 0
  (ALL x. P x) --> (ALL y. Q y --> (ALL z. P z))
   1. (ALL x. P x) --> (ALL y. Q y --> (ALL z. P z))
```

Proof checking can be tiresome, so let us save steps using repetition. We arrive at a state where ($\forall I$) and ($\rightarrow I$) have been applied as much as possible to subgoal 1:

```
by (REPEAT (resolve_tac [allI, impI] 1));
  Level 1
  (ALL x. P x) --> (ALL y. Q y --> (ALL z. P z))
   1. !!y z. [| ALL x. P x; Q y |] ==> P z
```

Subgoal 1 asks us to prove $Pz$ in a context consisting of arbitrary $y$ and $z$, and the assumptions $\forall x\, Px$ and $Qy$. The management of such contexts is the sense in which Horn clause resolution has been generalized. The treatment of quantifiers is a natural alternative to Skolemization [16].

The operation `eresolve_tac` applies rules to any suitable assumptions. Its effect here, with ($\forall E$), is to strip the quantifier:

```
by (eresolve_tac [allE] 1);
  Level 2
  (ALL x. P x) --> (ALL y. Q y --> (ALL z. P z))
   1. !!y z. [| Q y; P (?x4 y z) |] ==> P z
```

This proof state has a logical variable, `?x4`. It is a function variable, but you may prefer to think of `?x4 y z` as a placeholder for any term possibly involving $y$ and $z$.

```
by (assume_tac 1);
  Level 3
  (ALL x. P x) --> (ALL y. Q y --> (ALL z. P z))
  No subgoals!
```

Proof by assumption instantiates the placeholder to just $z$, proving the theorem. Occurrences of logical variables elsewhere in the goal clause are updated as one would expect: Isabelle can be used for answer extraction. Few proof checkers allow such placeholders.

## 5 Beyond Proof Checking

A generic environment for proof checking is all very well, but what about automation? Functions like `resolve_tac`, which operate on the proof state, are called *tactics*. Functions like `REPEAT`, which operate on tactics, are called *tacticals*. Tactics and tacticals were conceived by Robin Milner and became popular through the HOL system. They constitute a language for expressing proofs at a high level. Isabelle supports automation through a novel treatment of tactics and tacticals, using ideas from logic programming.

Combining the typed $\lambda$-calculus with unification requires higher-order unification [9]. Above we saw the unification of $\lambda yz.P(?x_4\ y\ z)$ with $\lambda yz.P\ z$, yielding the unifier $?x_4 \mapsto \lambda yz.z$. A pair of terms may have infinitely many higher-order unifiers. Accordingly, all Isabelle tactics allow an inference to generate a lazy list of outcomes. Multiple unifiers are seldom useful in practice; restrictions of higher-order unification, such as pattern unification [15, 19], are easier to implement and still provide an adequate basis for Isabelle's approach to proof checking.

Multiple unifiers or not, we need lazy lists. They support backtracking, letting Isabelle supply the basic components of search engines — in the form of tacticals. In addition to `REPEAT` for repetition, there are tacticals `DEPTH_FIRST`, `BEST_FIRST`, `ITER_DEEPEN`, etc., for the usual search strategies. Other tacticals, like Prolog's cut, suppress alternatives. Expert users have fine control over the search space and how it is searched.

As one application I have implemented model elimination, the method underlying SETHEO and PTTP. But it does not harmonize with the generic spirit of Isabelle. Working in an applied theory requires unfolding all definitions until the problem is reduced to pure logic, or supplying additional properties as axioms.

More useful in proofs are the simple tools of Isabelle's classical reasoner. They are implemented using tacticals. They accept and use inference rules derived in the user's domain, just as a rewriter accepts any rewrite rules. The classical reasoner applies unification in a nonclausal setting: deductive tableaux.

## 6 Theorem Proving with Tableaux

The tableau method has much in common with resolution. There are tableau provers of great sophistication, like HARP [22]. But leanTAP works surprisingly well in spite of extreme simplicity [1]. It puts the negated conjecture into negation normal form; after Skolemization, the only remaining logical constants are $\land$, $\lor$ and $\forall$. A branch in the tableau is transformed in the usual way: $\phi \land \psi$ is replaced by the two formulæ $\phi$ and $\psi$, while $\phi \lor \psi$ splits the branch into two branches containing $\phi$ and $\psi$, respectively. The quantified formula $\forall x\, \phi$ augments the branch with a new instance of $\phi$. A branch is closed (deemed proven) if it contains a pair of complementary, unifiable literals; the unifier must be applied to the remaining branches.

At first sight, the tableau method may seem no more generic than resolution. But the logical constants, and corresponding branch operations, need not be built into the program. This point becomes clearer if we speak not of tableaux but in the equivalent language of sequents. The branch operations mentioned above are familiar rules of the sequent calculus:

$$\frac{\phi, \psi, \Gamma \Rightarrow \Delta}{\phi \wedge \psi, \Gamma \Rightarrow \Delta} \qquad \frac{\phi, \Gamma \Rightarrow \Delta \quad \psi, \Gamma \Rightarrow \Delta}{\phi \vee \psi, \Gamma \Rightarrow \Delta} \qquad \frac{\phi[t/x], \forall x\, \phi, \Gamma \Rightarrow \Delta}{\forall x\, \phi, \Gamma \Rightarrow \Delta}$$

Such rules can be given declaratively; we can add new rules at any time. Here is a rule for intersections:

$$\frac{x \in A, x \in B, \Gamma \Rightarrow \Delta}{x \in A \cap B, \Gamma \Rightarrow \Delta}$$

It describes the branch operation of replacing $x \in A \cap B$ by the two formulæ $x \in A$ and $x \in B$. This may appear to be of little value: one could just rewrite by the equivalence

$$x \in A \cap B \iff x \in A \wedge x \in B$$

to eliminate occurrences of $x \in A \cap B$. But occurrences of $A \cap B$ can be created during the proof, so this rewriting would have to be interleaved with tableau operations (where it might be called demodulation).

A resolution prover could, of course, take the equivalence above as an axiom. Three clauses would result:

$$\begin{aligned} x \in A \cap B &\leftarrow x \in A, x \in B \\ x \in A &\leftarrow x \in A \cap B \\ x \in B &\leftarrow x \in A \cap B \end{aligned}$$

Model elimination needs all contrapositives of these, too, making a total of seven clauses. Adding similar axioms for $A \cup B$, $A - B$, etc., makes the search space explode. But the sequent/tableau approach handles the set operations easily.

The sequent rules above all act on the left of the arrow, $\Rightarrow$. The sequent calculus also has rules that act on the right. leanTAP does not need them because it starts with a sequent of the form $\phi \Rightarrow \mathbf{false}$; since $\phi$ is in negation normal form, it contains no connectives that would move a subformula of $\phi$ to the right. But normal forms are unreadable and hard to reconcile with generic reasoning: the normalizer would have to be extended dynamically to cope with new operations. So we also need rules that act on the right:

$$\frac{\Gamma \Rightarrow \Delta, \phi, \psi}{\Gamma \Rightarrow \Delta, \phi \vee \psi} \qquad \frac{\Gamma \Rightarrow \Delta, x \in A \quad \Gamma \Rightarrow \Delta, x \in B}{\Gamma \Rightarrow \Delta, x \in A \cap B}$$

# 7  Tableau Methods in Isabelle

Sequent calculus rules can be represented easily in Isabelle. One of Isabelle's object-logics is Gentzen's calculus LK. An established technique [10] reduces associative unification to higher-order unification; Isabelle/LK directly reasons about sequents of the form

$$\phi_1, \ldots, \phi_m \Rightarrow \psi_1, \ldots, \psi_n$$

using rules that look almost as they would in standard texts [35].

However, natural deduction seems easier to automate than the sequent calculus. It works with formulæ instead of sequents. We do not need associative unification; Isabelle has built-in support for its manipulation of assumptions. The correspondence between natural deduction and sequent calculus rules is simple. Instead of rules that act on the right of the $\Rightarrow$ symbol, we have introduction rules like

$$\frac{x \in A \quad x \in B}{x \in A \cap B.}$$

Instead of rules that act on the left, we have elimination rules:

$$\frac{x \in A \cap B \quad \begin{array}{c}[x \in A,\ x \in B]\\ \psi\end{array}}{\psi.}$$

Both rules replace some occurrence of $x \in A \cap B$ by separate occurrences of $x \in A$ and $x \in B$.

Natural deduction rules do not refer to sets of formulæ, such as $\Gamma$ and $\Delta$: they are implicit. One vestige of the sequent style is the formula $\psi$ mentioned in the $\cap$-elimination rule above. Isabelle goal clauses contain sequents of the form $\Gamma \Rightarrow \phi$. Isabelle's standard treatment of natural deduction rules handles the usual operations on $\Gamma$. It can even delete assumptions (using `eresolve_tac`), which is seldom seen in natural deduction but is needed for the sequent calculus. An obvious use of contrapositives yields the effect of having multiple formulæ to the right of the $\Rightarrow$ symbol. The documentation provides further details [27].

For instance, the natural deduction form of the rule
$$\frac{x \in A, x \in B, \Gamma \Rightarrow \Delta}{x \in A \cap B, \Gamma \Rightarrow \Delta}$$
is the $\cap$-elimination rule shown above, whose Isabelle form is the nested Horn clause
$$[\![ x \in A \cap B; [\![ x \in A; x \in B ]\!] \Rightarrow \phi ]\!] \Rightarrow \phi.$$

Isabelle's classical reasoner supplies several sequent-style tactics. Each accepts a collection of sequent rules, packaged into a *classical set* or *claset*. Each rule is augmented with hints concerning its use. It is specified as introduction or elimination and as *safe* or *unsafe*. A rule is safe if its premises are logically equivalent to its conclusion, while each somehow "smaller" (to ensure termination). Safe rules can be applied at any time; unsafe rules risk a loss of completeness and are applied only as a last resort. These two rules are unsafe:
$$\frac{\phi, \Gamma \Rightarrow \Delta}{\forall x\, \phi, \Gamma \Rightarrow \Delta} \qquad \frac{\Gamma \Rightarrow \Delta, x \in A \quad x \in B, \Gamma \Rightarrow \Delta}{A \subseteq B, \Gamma \Rightarrow \Delta}$$
The quantifier rule, unlike the conventional one given previously, discards each $\forall$-formula after a single use. The subset rule similarly discards a formula of the form $A \subseteq B$. Such rules let us adopt a simple depth-first search procedure without looping. This treatment of quantifiers is grossly incomplete: it cannot even prove $\exists x \forall y\, (\psi x \rightarrow \psi y)$. But such formulæ seldom arise in practice.

The classical reasoning tactic `Fast_tac` performs depth-first search. It is one of Isabelle's most commonly used tactics. There is also `Best_tac`, which uses the same rules under best-first search.[2]

For example, consider problem 38 of Pelletier [31]. One of the harder ones, it poses difficulties for leanTAP [1]. Like many problems that arise in practice, it is long but logically shallow:

---

[2] Users may bind specialized clasets to ML identifiers, but Isabelle now supports a *default claset*. Tactics `Fast_tac`, `Best_tac`... refer to the default claset, while `fast_tac`, `best_tac`... take a claset as argument. For simplicity I discuss only the capitalized versions here.

```
goal FOL.thy
    "(ALL x. p(a) & (p(x) --> (EX y. p(y) & r(x,y))) -->    \
\            (EX z. EX w. p(z) & r(x,w) & r(w,z))) <->      \
\    (ALL x. (~p(a) | p(x) | (EX z. EX w. p(z) & r(x,w) & r(w,z))) &    \
\            (~p(a) | ~(EX y. p(y) & r(x,y)) |              \
\             (EX z. EX w. p(z) & r(x,w) & r(w,z))))";
```

It does not require quantifier duplication, so `Fast_tac` succeeds, yielding a 153-step proof in about four seconds.[3] Most of Pelletier's problems take a second or less.

What good is a tool that needs seconds, not milliseconds, to solve problems that by modern standards are trivial? The tactic can also solve problems that cannot even be expressed easily in other systems. Consider the following identity of set theory:

$$\bigcup_{i \in I}(A_i \cup B_i) = \Big(\bigcup_{i \in I} A_i\Big) \cup \Big(\bigcup_{i \in I} B_i\Big)$$

We can supply this literally to Isabelle and prove it using `Fast_tac`:

```
goal Set.thy "(UN i:I. A(i) Un B(i)) = (UN i:I. A(i))  Un  (UN i:I. B(i))";
```

This takes about 0.3 seconds, yielding a 27-step proof. Thanks to the use of rules about set theory, the internal proof tree is concise. Reasoning takes place on the level of unions and intersections, which are not expanded to their definitions. The result is greater efficiency, vital when the underlying proof engine can perform only a few hundred inferences per second. Moreover, proofs (and failed proofs) become easier to understand.

We do not have to keep the set of rules small. For this and many other proofs, the search space is tiny. Backtracking is minimized through the notion of safe proof step. A typical safe step applies a safe rule without instantiating logical variables. Let us contrast safe and unsafe uses of the rule

$$[\![ x \in A; x \in B ]\!] \Longrightarrow x \in A \cap B.$$

Splitting the goal $?w \in R \cap S$ into the subgoals $?w \in R$ and $?w \in S$ is safe, since the logical variable $?w$ stays unchanged. On the other hand, splitting the goal $w \in ?R$ into the subgoals $w \in ?R_1$ and $w \in ?R_2$ instantiates $?R$ to $?R_1 \cap ?R_2$. It is unsafe for two reasons: (1) it might falsify other subgoals, and (2) such steps can be repeated forever. They amount to guessing; backtracking can usefully consider other instantiations of $?R$.

---

[3]Timings were conducted on a Sun SuperSPARC Model 61.

Occasionally, the search space does explode. Proving Cantor's theorem requires guessing a term composed of several set operations. It defeats `Fast_tac` entirely. It can be solved using `Best_tac`, but the claset must kept minimal [25, §8.2]. This famous theorem requires creativity and we should expect its proof to be difficult.

Rudnicki [33] attributes to Martin Davis a definition of *obvious* theorems: those that can be proved using each universal premise at most once. Under this definition `Fast_tac` can only prove obvious theorems. But this notion is relative to the set of rules supplied in the claset. The set-theoretic identity proved above is obvious with respect to high-level rules about sets, but perhaps not if it is reduced to pure first-order logic. Supplied rules can be instantiated any number of times.[4]

These tools are not merely applicable to first-order logic and set theory. They are indeed generic, with a variety of applications. To illustrate this, let us consider a substantial proof development.

## 8 The Church-Rosser Theorem for Combinators

The language of combinators has two constants, **S** and **K**, and an apply operation written by juxtaposition; if $x$ and $y$ are terms then $xy$ is the term for $x$ applied to $y$. Application associates to the left: $xyz$ abbreviates $(xy)z$. The combinators satisfy the *contractions*

$$\begin{aligned} \mathbf{K}xy &\stackrel{1}{\longrightarrow} x \\ \mathbf{S}xyz &\stackrel{1}{\longrightarrow} xz(yz). \end{aligned}$$

We also have $x \stackrel{1}{\longrightarrow} y$ if one of the contractions above can be applied to a subterm of $x$. An *inductive definition* expresses all this formally and concisely (Figure 2).

The only legal contractions are those shown in the figure. No matter what $z$ is, $\mathbf{K} \stackrel{1}{\longrightarrow} z$ is impossible: none of the rules can yield a conclusion of that form. Similarly, $\mathbf{S} \stackrel{1}{\longrightarrow} z$ is impossible. The only contractions for $\mathbf{K}x$ have the form $\mathbf{K}x \stackrel{1}{\longrightarrow} \mathbf{K}x'$ where $x \stackrel{1}{\longrightarrow} x'$. Analogous results hold for $\mathbf{S}x$ and $\mathbf{S}xy$. This sort of case analysis, examining the rules of an inductive definition to see what is possible, is called *rule inversion*.

---

[4]Rudnicki augments the single-use restriction with conditions to minimize search. By this stronger criterion, `Fast_tac` can prove non-obvious theorems.

$$\overline{\mathbf{K}xy \xrightarrow{1} x} \quad \overline{\mathbf{S}xyz \xrightarrow{1} xz(yz)}$$

$$\frac{x \xrightarrow{1} y}{xz \xrightarrow{1} yz} \quad \frac{x \xrightarrow{1} y}{zx \xrightarrow{1} zy}$$

Figure 2: Inductive Definition of Contraction

The identity combinator $\mathbf{I}$ is defined by $\mathbf{I} \equiv \mathbf{SKK}$:

$$\mathbf{SKK}x \xrightarrow{1} \mathbf{K}x(\mathbf{K}x) \xrightarrow{1} x.$$

No contraction of the form $\mathbf{I} \xrightarrow{1} z$ is possible. The only contractions for $\mathbf{SKK}$ have the form $\mathbf{SKK} \xrightarrow{1} \mathbf{S}xy$ where $\mathbf{K} \xrightarrow{1} x$ and $\mathbf{K} \xrightarrow{1} y$, and we have already seen that $\mathbf{K}$ does not contract to anything. This reasoning involves two levels of rule inversion.

*Rule induction* [36] is a more powerful inference rule for proving consequences of $x \xrightarrow{1} y$. Defining a relation inductively specifies the least set closed under the given rules. Formally, it is the least fixedpoint of a function over sets. While rule inversion is sound for any fixedpoint, rule induction is sound only for the least fixedpoint. To prove that $x \xrightarrow{1} y$ implies $\psi(x, y)$ for all $x$ and $y$, show that each of the rules defining $\xrightarrow{1}$ preserves $\psi$:

- $\psi(\mathbf{K}xy, x)$

- $\psi(\mathbf{S}xyz, xz(yz))$

- if $\psi(x, y)$ and $x \xrightarrow{1} y$ then $\psi(xz, yz)$

- if $\psi(x, y)$ and $x \xrightarrow{1} y$ then $\psi(zx, zy)$

The last two cases have $\psi(x, y)$ as an inductive hypothesis. Rule induction gives the effect of induction on the size of a proof, without tiresome arithmetic reasoning. Church-Rosser properties provide good examples for demonstrating it.

Say $x$ *reduces* to $y$, written $x \longrightarrow y$, if a series of contractions takes $x$ to $y$. Formally, reduction is the reflexive/transitive closure of contraction. Reduction satisfies the Church-Rosser property, which follows easily from the *diamond property*: if $x \longrightarrow y$ and $x \longrightarrow y'$ then there exists $z$ such that $y \longrightarrow z$ and $y' \longrightarrow z$:

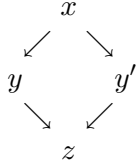

Camilleri and Melham [2, §4] discuss a proof of the diamond property to illustrate inductive definitions in HOL. They wrote ML code to automate parts of the proof, but still the proof script contains long chains of low-level inferences. The Isabelle proof script is simple and short, thanks to `Fast_tac`.[5]

## 9 Reasoning about Contraction in Isabelle

Figure 3 presents the Isabelle/HOL definitions of combinators and combinator reduction. Isabelle converts the inductive definition into a least fixedpoint definition from which it derives the specified rules [26]. It also derives rules for induction and case analysis. It returns an ML function, `contract.mk_cases`, for performing one level of rule inversion. To apply it to the combinator **K**, we type a `val` declaration at the ML top level:

```
val K_contractE = contract.mk_cases comb.simps "K -1-> z";
  val K_contractE = "K -1-> ?z ==> ?Q" : thm
```

The ML identifier `K_contractE` is now bound to the derived rule $\dfrac{\mathbf{K} \stackrel{1}{\longrightarrow} z}{Q}$.

The rule expresses the impossibility of $\mathbf{K} \stackrel{1}{\longrightarrow} z$ by stating that such a contraction implies anything. Isabelle (including `Fast_tac`) handles assertions of this form as elimination rules. Two further `val` declarations derive similar rules for the combinator **S** and for application:

```
val S_contractE = contract.mk_cases comb.simps "S -1-> z";
  val S_contractE = "S -1-> ?z ==> ?Q" : thm
val Ap_contractE = contract.mk_cases comb.simps "x#y -1-> z";
```

---

[5]Comparing the HOL and Isabelle developments is instructive. To obtain either system, consult the URL http://www.cl.cam.ac.uk/Research/HVG/. The HOL version is distributed in the **contrib** library in directory **rule-induction**. The Isabelle/HOL version is the theory `HOL/ex/Comb`; an earlier version for Isabelle/ZF is the theory `ZF/ex/Comb`.

```
datatype comb = K
              | S
              | "#" comb comb (infixl 90)     (*infix application operator*)

(** Inductive definition of contractions, -1->
            and (multi-step) reductions, --->   **)
consts
  contract  :: "(comb*comb) set"
  "-1->"    :: [comb,comb] => bool    (infixl 50)
  "--->"    :: [comb,comb] => bool    (infixl 50)

translations
  "x -1-> y" == "(x,y) : contract"
  "x ---> y" == "(x,y) : contract^*"

inductive "contract"
  intrs
    K      "K#x#y -1-> x"
    S      "S#x#y#z -1-> (x#z)#(y#z)"
    Ap1    "x-1->y ==> x#z -1-> y#z"
    Ap2    "x-1->y ==> z#x -1-> z#y"
```

Figure 3: Isabelle/HOL Definitions of Combinators, Contraction and Reduction

```
    val Ap_contractE =
      "[| ?x # ?y -1-> ?z;
          ?x = K # ?z ==> ?Q;
          !!x y. [| ?z = x # ?y # (y # ?y); ?x = S # x # y |] ==> ?Q;
          !!y. [| ?x -1-> y; ?z = y # ?y |] ==> ?Q;
          !!y. [| ?y -1-> y; ?z = ?x # y |] ==> ?Q |] ==> ?Q" : thm
```

Note that `Ap_contractE` splits $xy \xrightarrow{1} z$ into four cases.[6] The first case is $x = \mathbf{K}z$, where the contraction has the form $\mathbf{K}zy \xrightarrow{1} z$. There is a case for each of the four rules of the inductive definition (Figure 2) because each conclusion involves an application.

The function `contract.mk_cases` cannot prove the impossibility of $\mathbf{I} \xrightarrow{1} z$ because it requires case analysis to two levels:

```
contract.mk_cases (I_def::comb.simps) "I -1-> z";
  "[| S # K # K -1-> ?z;
      !!y. [| S # K -1-> y; ?z = y # K |] ==> ?Q;
      !!y. [| K -1-> y; ?z = S # K # y |] ==> ?Q |] ==> ?Q" : thm
```

The result is correct but not in simplest form; it splits its premise into two cases, both of which contain contradictory assumptions. We can get around this limitation of `mk_cases` by using Fast_tac to implement an effective search strategy.

Underlying `contract.mk_cases` is an elimination rule for contraction. This rule splits any assumption of the form $\ldots \xrightarrow{1} \ldots$ into four cases, one for each of the contraction rules. Two of these cases have new assumptions of the form $\ldots \xrightarrow{1} \ldots$, so the rule could be applied forever. In contrast, the rules `K_contractE`, `S_contractE` and `Ap_contractE` apply to assumptions of the form $\mathbf{K} \xrightarrow{1} z$, $\mathbf{S} \xrightarrow{1} z$ and $xy \xrightarrow{1} z$, respectively. The first two prove the goal outright; the third breaks up $xy \xrightarrow{1} z$, making smaller contractions of the form $x \xrightarrow{1} w$ or $y \xrightarrow{1} w$. These rules can be repeated safely in depth-first search. If initially we have an assumption of the form $x \xrightarrow{1} y$, they break up $x$ to atoms.

Let us add some rules to the default claset:

```
AddSEs [K_contractE, S_contractE, Ap_contractE];
AddIs  contract.intrs;
Addss  (!simpset);
```

---

[6]Strictly speaking, this is $?x?y \xrightarrow{1} ?z$; I omit the question marks for clarity. Note that ?x and x are distinct variables.

The `AddSEs` command stores the three `contract` rules as safe elimination rules. The `AddIs` command stores the rules of Figure 2 as introduction rules; they are unsafe because many terms can be contracted in more than one way. The `Addss` command asks the classical reasoner to perform demodulation (rewriting) using the default set of rewrite rules. They include the freeness properties of the combinators, which can replace **K** = **S** by **false** and **K**$x$ = **K**$y$ by $x = y$.

Now we can prove the impossibility of $\mathbf{I} \xrightarrow{1} z$. After we state the goal and replace **I** by **SKK**, a call to `Fast_tac` succeeds in a fraction of a second:

```
goalw Comb.thy [I_def] "!!z. I -1-> z ==> P";
  Level 0
  !!z. I -1-> z ==> P
   1. !!z. S # K # K -1-> z ==> P
by (Fast_tac 1);
  No subgoals!
```

## 10   Parallel Contraction

The reasoning style shown above handles most of the Church-Rosser development. The next stage is to define *parallel contraction*, $x \xRightarrow{1} y$ inductively:

$$\overline{x \xRightarrow{1} x} \qquad \overline{\mathbf{K}xy \xRightarrow{1} x}$$

$$\overline{\mathbf{S}xyz \xRightarrow{1} xz(yz)} \qquad \frac{x \xRightarrow{1} y \quad z \xRightarrow{1} w}{xz \xRightarrow{1} yw}$$

By rule induction over this set of rules, we can prove that $\xRightarrow{1}$ satisfies the diamond property. Figure 4 presents the necessary Isabelle definitions.

Proofs about parallel contraction involve more cases than those about ordinary contraction. For example, $K \xRightarrow{1} x$ is possible; we have $K \xRightarrow{1} K$. But the rule inversion techniques outlined above carry us to the proof that $\xRightarrow{1}$ satisfies the diamond property.

The ML function `parcontract.mk_cases` performs rule inversion with respect to parallel contraction. Let us apply it to the three forms of combinator terms: **K**, **S** and application.

```
val K_parcontractE = parcontract.mk_cases comb.simps "K =1=> z";
  val K_parcontractE = "[| K =1=> ?z; ?z = K ==> ?Q |] ==> ?Q" : thm
```

```
(** Inductive definition of parallel contractions, =1=>
           and (multi-step) parallel reductions, ===>  **)
consts
  parcontract :: "(comb*comb) set"
  "=1=>"      :: [comb,comb] => bool    (infixl 50)
  "===>"      :: [comb,comb] => bool    (infixl 50)

translations
  "x =1=> y" == "(x,y) : parcontract"
  "x ===> y" == "(x,y) : parcontract^*"

inductive "parcontract"
  intrs
    refl  "x =1=> x"
    K     "K#x#y =1=> x"
    S     "S#x#y#z =1=> (x#z)#(y#z)"
    Ap    "[| x=1=>y;  z=1=>w |] ==> x#z =1=> y#w"

constdefs
  diamond   :: "('a * 'a)set => bool"
  "diamond(r) == ALL x y. (x,y):r -->
                 (ALL y'. (x,y'):r -->
                   (EX z. (y,z):r & (y',z) : r))"
```

Figure 4: Isabelle/HOL Definitions of Parallel Contraction and Reduction and the Diamond Property

```
val S_parcontractE = parcontract.mk_cases comb.simps "S =1=> z";
  val S_parcontractE = "[| S =1=> ?z; ?z = S ==> ?Q |] ==> ?Q" : thm
val Ap_parcontractE = parcontract.mk_cases comb.simps "x#y =1=> z";
  val Ap_parcontractE = "[| ?x # ?y =1=> ?z; ... |] ==> ?Q" : thm
```

We can add these rules to the default claset, as we added the `contract` rules above. Case analysis rules for terms of the form $\mathbf{K}x$, $\mathbf{S}x$ and $\mathbf{S}xy$ can then be proved trivially, using `Fast_tac`. Here is one of them, which expresses that all parallel contractions on $\mathbf{K}x$ have the form $\mathbf{K}x \stackrel{1}{\Longrightarrow} \mathbf{K}x'$, where $x \stackrel{1}{\Longrightarrow} x'$.

```
goal Comb.thy "!!x z. K#x =1=> z ==> (EX x'. z = K#x' & x =1=> x')";
  Level 0
  !!x z. K # x =1=> z ==> EX x'. z = K # x' & x =1=> x'
   1. !!x z. K # x =1=> z ==> EX x'. z = K # x' & x =1=> x'
by (Fast_tac 1);
  No subgoals!
```

We name this theorem and add it to the default claset, so that it can speed future proofs.

```
val K1_parcontractD = result();
  val K1_parcontractD = "K # ?x =1=> ?z ==> EX x'. ?z = K # x' & ..." : thm
AddSDs [K1_parcontractD];
```

The diamond property for $\stackrel{1}{\Longrightarrow}$ has a three-step proof. A first, trivial step yields this proof state:

```
  Level 1
  diamond parcontract
   1. !!y xa. y =1=> xa ==> ALL y'. y =1=> y' -->
                                  (EX z. xa =1=> z & y' =1=> z)
```

We apply rule induction to the assumption, $y \stackrel{1}{\Longrightarrow} xa$. Rule induction involves showing that the quantified formula above is preserved over the four rules that define parallel contraction. Figure 5 presents the four cases to be proved. The first subgoal is the case $xa = y$, and is easily proved: if $x \stackrel{1}{\Longrightarrow} y'$ then choose $z = y'$; we get $y' \stackrel{1}{\Longrightarrow} y'$ by reflexivity.

The final step applies `Fast_tac` to all four cases.

```
by (ALLGOALS Fast_tac);
  Level 3
  diamond parcontract
  No subgoals!
```

```
Level 2
diamond parcontract
 1. !!x. ALL y'. x =1=> y' --> (EX z. x =1=> z & y' =1=> z)
 2. !!x ya. ALL y'. K # x # ya =1=> y' --> (EX z. x =1=> z & y' =1=> z)
 3. !!x ya z.
       ALL y'.
          S # x # ya # z =1=> y' -->
          (EX za. x # z # (ya # z) =1=> za & y' =1=> za)
 4. !!w x ya z.
       [| x =1=> ya; ALL y'. x =1=> y' --> (EX z. ya =1=> z & y' =1=> z);
          z =1=> w;  ALL y'. z =1=> y' --> (EX z. w =1=> z & y' =1=> z) |] ==>
       ALL y'. x # z =1=> y' --> (EX z. ya # w =1=> z & y' =1=> z)
```

Figure 5: Four Cases of Induction for Diamond Property

Consider how we might prove these cases interactively. Users typically apply `Fast_tac` to one goal at a time. The first subgoal is proved in a fraction of a second; the next one takes about a second; the third takes about three seconds. But the fourth subgoal, the inductive step, takes over forty seconds; a user might interrupt the proof attempt.

Fortunately, one can single-step `Fast_tac`'s proof strategy. The tactic `safe_tac` uses a given claset to perform only safe inferences on the proof state. It quickly breaks up complicated subgoals into parts that can be proved independently. It splits our inductive step into four new subgoals (Figure 6). They arise because the formula `x#z =1=> y'` becomes an assumption, which `Ap_parcontractE` breaks up.

Continuing the interactive proof, our user would find that `Fast_tac` solves the first three subgoals in a few seconds each. The fourth subgoal still takes nearly thirty seconds, but perhaps this is not too much for human patience.

Applying `safe_tac` to the original four subgoals of the induction (Figure 5) produces a total of thirteen subgoals. We see that proving the diamond property requires a substantial case analysis. Such proofs are tiresome for humans but easy for `Fast_tac`.

So parallel contraction ($\overset{1}{\Longrightarrow}$) satisfies the diamond property. The diamond property for parallel reduction, $\Longrightarrow$, follows from a general theorem about the reflexive/transitive closure: $\text{diamond}(R)$ implies $\text{diamond}(R^*)$. It is proved by two nested inductions, whose subcases are proved using

```
Level 6
diamond parcontract
 1. !!w x ya z y'.
       [| x =1=> ya;
          ALL y'. x =1=> y' --> (EX z. ya =1=> z & y' =1=> z); z =1=> w;
          ALL y'. z =1=> y' --> (EX z. w =1=> z & y' =1=> z) |] ==>
       EX za. ya # w =1=> za & x # z =1=> za
 2. !!w x ya z y' x'.
       [| ALL y'a.
              K # y' =1=> y'a --> (EX z. K # x' =1=> z & y'a =1=> z);
          z =1=> w; ALL y'. z =1=> y' --> (EX z. w =1=> z & y' =1=> z);
          y' =1=> x' |] ==>
       EX z. K # x' # w =1=> z & y' =1=> z
 3. !!w x ya z y' xa y x' y'a.
       [| ALL y'.
              S # xa # y =1=> y' -->
              (EX z. S # x' # y'a =1=> z & y' =1=> z);
          z =1=> w; ALL y'. z =1=> y' --> (EX z. w =1=> z & y' =1=> z);
          xa =1=> x'; y =1=> y'a |] ==>
       EX za. S # x' # y'a # w =1=> za & xa # z # (y # z) =1=> za
 4. !!w x ya z y' wa y.
       [| x =1=> ya;
          ALL y'. x =1=> y' --> (EX z. ya =1=> z & y' =1=> z); z =1=> w;
          ALL y'. z =1=> y' --> (EX z. w =1=> z & y' =1=> z); x =1=> y;
          z =1=> wa |] ==>
       EX z. ya # w =1=> z & y # wa =1=> z
```

Figure 6: Four Subcases of the Inductive Step

Fast_tac and a variant of Best_tac.[7]

Our objective is to show that reduction ($\longrightarrow$) satisfies the diamond property. We can do this by showing that $\Longrightarrow$ is equivalent to $\longrightarrow$.

## 11  The Equivalence of $\Longrightarrow$ and $\longrightarrow$

Proving the equivalence of $x \Longrightarrow y$ and $x \longrightarrow y$ involves several steps. We can state it as the set equality $(\overset{1}{\Longrightarrow})^* = (\overset{1}{\longrightarrow})^*$ and prove the two inclusions separately. The proof of $(\overset{1}{\longrightarrow}) \subseteq (\overset{1}{\Longrightarrow})$ is trivial using the methods already discussed. Then $(\overset{1}{\longrightarrow})^* \subseteq (\overset{1}{\Longrightarrow})^*$ follows by the monotonicity of the reflexive/transitive closure.

The opposite inclusion is $(\overset{1}{\Longrightarrow})^* \subseteq (\overset{1}{\longrightarrow})^*$. By monotonicity again, this reduces to $(\overset{1}{\Longrightarrow}) \subseteq (\overset{1}{\longrightarrow})^*$. Rule induction on $\Longrightarrow$ leaves four subgoals:

```
by (etac parcontract.induct 1);
  Level 3
  parcontract <= contract^*
   1. !!x. x ---> x
   2. !!x ya. K # x # ya ---> x
   3. !!x ya z. S # x # ya # z ---> x # z # (ya # z)
   4. !!w x ya z.
        [| x =1=> ya; x ---> ya; z =1=> w; z ---> w |] ==>
        x # z ---> ya # w
```

Each subgoal involves proving something of the form $x \longrightarrow y$: a series of contractions. We need rewriting by $\mathbf{K}xy \overset{1}{\longrightarrow} x$ and $\mathbf{S}xyz \overset{1}{\longrightarrow} xz(yz)$, with recursive rewriting of subterms. Isabelle's built-in simplifier allows rewriting only with respect to equality. Rewriting with respect to other transitive relations can be implemented by deriving rules that amount to a Prolog program [18], but that is too complicated for such a simple proof.

Instead we just combine the contractions for $\mathbf{K}$ and $\mathbf{S}$ with lemmas saying that $x \longrightarrow y$ implies $xz \longrightarrow yz$ and $zx \longrightarrow zy$. We add lemmas that $R^*$ includes $R$ and is transitive.

```
AddSIs [contract.K, contract.S];
AddIs  [Ap_reduce1, Ap_reduce2, r_into_rtrancl, rtrancl_trans];
```

We can no longer use Fast_tac: the transitive law would make the depth-first strategy loop. The classical reasoner provides a tactic that uses iterative deepening: Deepen_tac. It proves all four subgoals in one second.

---

[7]Camilleri and Melham [2, §4.4.2] explain the ideas behind this proof. Paulson [28, §2.5] discusses inductive proofs about the transitive closure.

This tactic takes ideas from `Fast_tac` and leanTAP. It retains the notion of safe inference and uses the same kinds of clasets. But in every unsafe inference it duplicates the affected formula. (This includes quantifier duplication as a special case.) The depth bound limits the number of unsafe inferences performed. Thus `Deepen_tac` is probably complete. It can prove several theorems that `Fast_tac` cannot manage, such as Pelletier's problems 18, 19, 21, 34 (Andrews's Challenge), 42, 43, 50 and 59. Outside first-order logic its performance is disappointing; for problems that `Fast_tac` cannot handle, `Deepen_tac` is often too slow. We have the usual trade-off between completeness and efficiency in proof procedures.

## 12 Conclusions

Many problems in programming language semantics concern reductions on complex terms: the evaluation of expressions, the execution of commands. Such problems can be defined inductively and reasoned about using the techniques described above. Nipkow [20] and Rasmussen [32] have developed independent proofs of the Church-Rosser theorem for the $\lambda$-calculus. Lötzbeyer, Sandner and Nipkow [12, 21] have proved several properties relating the operational and denotational semantics of Winskel's toy programming language IMP [36].

A *coinductive* definition is the dual of an inductive definition: a greatest fixedpoint instead of a least one. Rule inversion remains sound for coinductive definitions. Frost [6] has used it in a large development, formalizing work by Milner and Tofte [17]. The same combination of `Fast_tac` and `mk_cases` appears in several proofs.

The classical reasoner finds broader uses. This is not surprising: the concepts of safe and unsafe inference are intuitive and give control over backtracking. I have written a crude shell script to count occurrences of tactics in a set of files.[8] Over the whole of the Isabelle/ZF distribution there are 9200 tactic occurrences, of which 1500 refer to `Fast_tac`. This exceeds the number of calls to the rewriter (1200) but falls well short of the number of explicit proof checking steps. A total of 2800 theorems are proved; thus, proving a theorem requires about 3.3 tactic invocations on average.

In Isabelle/HOL there are 7300 tactic occurrences, including nearly 1100 `Fast_tac` calls and 1400 rewriter calls. With over 2200 theorems proved, a proof on average comprises 3.3 tactic calls — a remarkable agreement with ZF. The files examined comprise major developments by a variety of authors [6, 20, 30, 32]. Isabelle/HOL is largely the work of Tobias Nipkow, while Isabelle/ZF is largely my own work. This diversity is further evidence that the classical reasoner is generic.

---

[8]The figures have been rounded to two significant digits. The script counts only theorems proved using tactics, omitting `K_contractE`, for instance.

In the combinator reduction example, 20 theorems are proved using only 37 tactic calls, of which 18 are to `Fast_tac` or other classical reasoning tools. In contrast, Camilleri and Melham's original development invokes approximately 230 tactics and 30 conversions (rewriting primitives), hand-coded into specialized proof procedures.

There are two reasons for expecting the classical reasoner to perform well in other domains. First, consider the great diversity of the examples reported above, which range from computational examples to deep results in axiomatic set theory. Second, consider the simplicity of the tool itself, which is based upon using rules to break up formulæ, with the safe/unsafe criterion to control the search space.

A major goal in automated reasoning is to make our tools easy to use. The number of commands per theorem proved is only one measure of this. User interface improvements, such as automatic classification of rules, are still needed. Naïve tableau-based methods may be much weaker than classical resolution when applied to traditional challenge problems. In realistic proof developments, they are a valuable complement to decision procedures and rewriting tools.

**Acknowledgements.** J. Harrison and T. Nipkow made detailed comments on this article; M. Wenzel reported some errors in it. J. Camilleri and T. F. Melham's work [2] was an inspiration for my own; their lucid description of the Church-Rosser theorem for combinators was particularly helpful. The research was funded by the EPSRC GR/H40570 "Combining HOL and Isabelle" and by the ESPRIT Basic Research Action 6453 "Types."

# References

[1] B. Beckert and J. Posegga. leanTAP: Lean tableau-based deduction. *J. Auto. Reas.*, 15(3):339–358, 1995.

[2] J. Camilleri and T. F. Melham. Reasoning with inductively defined relations in the HOL theorem prover. Technical Report 265, Comp. Lab., Univ. Cambridge, Aug. 1992.

[3] D. Cyrluk, P. Lincoln, and N. Shankar. On Shostak's decision procedure for combinations of theories. In McRobbie and Slaney [14], pages 463–477.

[4] G. Dowek et al. The Coq proof assistant user's guide. Technical Report 154, INRIA-Rocquencourt, 1993. Version 5.8.

[5] A. Felty. Implementing tactics and tacticals in a higher-order logic programming language. *J. Auto. Reas.*, 11(1):43–82, 1993.

[6] J. Frost. A case study of co-induction in Isabelle. Technical Report 359, Comp. Lab., Univ. Cambridge, Feb. 1995.

[7] M. J. C. Gordon. Why higher-order logic is a good formalism for specifying and verifying hardware. In G. Milne and P. A. Subrahmanyam, editors, *Formal Aspects of VLSI Design*, pages 153–177. North-Holland, 1986.

[8] M. J. C. Gordon and T. F. Melham. *Introduction to HOL: A Theorem Proving Environment for Higher Order Logic.* Cambridge Univ. Press, 1993.

[9] G. P. Huet. A unification algorithm for typed $\lambda$-calculus. *Theoretical Comput. Sci.*, 1:27–57, 1975.

[10] G. P. Huet and B. Lang. Proving and applying program transformations expressed with second-order patterns. *Acta Inf.*, 11:31–55, 1978.

[11] R. Letz, J. Schumann, S. Bayerl, and W. Bibel. SETHEO: A high-performance theorem prover. *J. Auto. Reas.*, 8(2):183–212, 1992.

[12] H. Lötzbeyer and R. Sandner. Proof of the equivalence of the operational and denotational semantics of IMP in Isabelle/ZF. Project report, Institut für Informatik, TU München, 1994.

[13] W. McCune. OTTER 3.0 Reference Manual and Guide. Technical Report ANL-94/6, Argonne National Laboratory, Argonne, IL, 1994.

[14] M. McRobbie and J. K. Slaney, editors. *Automated Deduction — CADE-13 International Conference*, LNAI 1104. Springer, 1996.

[15] D. Miller. A logic programming language with lambda-abstraction, function variables, and simple unification. *J. Logic and Comput.*, 1(4):497–536, 1991.

[16] D. Miller. Unification under a mixed prefix. *J. Symb. Comput.*, 14(4):321–358, 1992.

[17] R. Milner and M. Tofte. Co-induction in relational semantics. *Theoretical Comput. Sci.*, 87:209–220, 1991.

[18] T. Nipkow. Constructive rewriting. *Comput. J.*, 34:34–41, 1991.

[19] T. Nipkow. Functional unification of higher-order patterns. In M. Vardi, editor, *Eighth Annual Symposium on Logic in Computer Science*, pages 64–74. IEEE Comp. Soc. Press, 1993.

[20] T. Nipkow. More Church-Rosser proofs (in Isabelle/HOL). In McRobbie and Slaney [14], pages 733–747.

[21] T. Nipkow. Winskel is (almost) right: Towards a mechanized semantics textbook. In *Foundations of Software Technology and Theoretical Computer Science*, LNCS. Springer, 1996. In press.

[22] F. Oppacher and E. Suen. HARP: A tableau-based theorem prover. *J. Auto. Reas.*, 4(1):69–100, 1988.

[23] S. Owre, J. M. Rushby, N. Shankar, and M. K. Srivas. A tutorial on using PVS for hardware verification. In R. Kumar, editor, *Theorem Provers in Circuit Design: Theory, Practice, and Experience*, LNCS 901, pages 258–279. Springer, 1995.

[24] L. C. Paulson. The foundation of a generic theorem prover. *J. Auto. Reas.*, 5(3):363–397, 1989.

[25] L. C. Paulson. Set theory for verification: I. From foundations to functions. *J. Auto. Reas.*, 11(3):353–389, 1993.

[26] L. C. Paulson. A fixedpoint approach to implementing (co)inductive definitions. In A. Bundy, editor, *Automated Deduction — CADE-12 International Conference*, LNAI 814, pages 148–161. Springer, 1994.

[27] L. C. Paulson. *Isabelle: A Generic Theorem Prover*. Springer, 1994. LNCS 828.

[28] L. C. Paulson. Set theory for verification: II. Induction and recursion. *J. Auto. Reas.*, 15(2):167–215, 1995.

[29] L. C. Paulson. *ML for the Working Programmer*. Cambridge Univ. Press, 2nd edition, 1996.

[30] L. C. Paulson and K. Grąbczewski. Mechanizing set theory: Cardinal arithmetic and the axiom of choice. *J. Auto. Reas.*, 1996. In press.

[31] F. J. Pelletier. Seventy-five problems for testing automatic theorem provers. *J. Auto. Reas.*, 2:191–216, 1986. Errata, JAR 4 (1988), 235–236.

[32] O. Rasmussen. The Church-Rosser theorem in Isabelle: A proof porting experiment. Technical Report 364, Computer Laboratory, University of Cambridge, May 1995.

[33] P. Rudnicki. Obvious inferences. *J. Auto. Reas.*, 3(4):383–393, 1987.

[34] M. E. Stickel. A Prolog technology theorem prover: Implementation by an extended Prolog compiler. *J. Auto. Reas.*, 4(4):353–380, 1988.

[35] G. Takeuti. *Proof Theory*. North-Holland, 2nd edition, 1987.

[36] G. Winskel. *The Formal Semantics of Programming Languages*. MIT Press, 1993.

# A   Full Proof Script

```
(*  Title:      HOL/ex/comb.ML
    ID:         $Id: Comb.ML,v 1.8 1996/08/19 09:18:36 paulson Exp $
    Author:     Lawrence C Paulson
    Copyright   1996  University of Cambridge

Combinatory Logic example: the Church-Rosser Theorem
Curiously, combinators do not include free variables.

Example taken from
    J. Camilleri and T. F. Melham.
    Reasoning with Inductively Defined Relations in the HOL Theorem Prover.
    Report 265, University of Cambridge Computer Laboratory, 1992.

HOL system proofs may be found in
```

```
/usr/groups/theory/hvg-aftp/contrib/rule-induction/cl.ml
*)

open Comb;

(*** Reflexive/Transitive closure preserves the Church-Rosser property
     So does the Transitive closure; use r_into_trancl instead of rtrancl_refl
***)

val [_, spec_mp] = [spec] RL [mp];

(*Strip lemma.  The induction hyp is all but the last diamond of the strip.*)
goalw Comb.thy [diamond_def]
    "!!r. [| diamond(r);  (x,y):r^* |] ==> \
\         ALL y'. (x,y'):r --> (EX z. (y',z): r^* & (y,z): r)";
by (etac rtrancl_induct 1);
by (Fast_tac 1);
by (slow_best_tac (!claset addIs [r_into_rtrancl RSN (2, rtrancl_trans)]
                           addSDs [spec_mp]) 1);
val diamond_strip_lemmaE = result() RS spec RS mp RS exE;

val [major] = goal Comb.thy "diamond(r) ==> diamond(r^*)";
by (rewtac diamond_def);  (*unfold only in goal, not in premise!*)
by (rtac (impI RS allI RS allI) 1);
by (etac rtrancl_induct 1);
by (Fast_tac 1);
by (ALLGOALS  (*Seems to be a brittle, undirected search*)
    (slow_best_tac ((claset_of "Fun") addIs [r_into_rtrancl, rtrancl_trans]
                           addEs [major RS diamond_strip_lemmaE])));
qed "diamond_rtrancl";

(*** Results about Contraction ***)

(** Non-contraction results **)

(*Derive a case for each combinator constructor*)
val K_contractE = contract.mk_cases comb.simps "K -1-> z";
val S_contractE = contract.mk_cases comb.simps "S -1-> z";
val Ap_contractE = contract.mk_cases comb.simps "x#y -1-> z";

AddIs   contract.intrs;
AddSEs [K_contractE, S_contractE, Ap_contractE];
Addss  (!simpset);

goalw Comb.thy [I_def] "!!z. I -1-> z ==> P";
by (Fast_tac 1);
```

```
qed "I_contract_E";
AddSEs [I_contract_E];

goal Comb.thy "!!x z. K#x -1-> z ==> (EX x'. z = K#x' & x -1-> x')";
by (Fast_tac 1);
qed "K1_contractD";
AddSEs [K1_contractD];

goal Comb.thy "!!x z. x ---> y ==> x#z ---> y#z";
by (etac rtrancl_induct 1);
by (ALLGOALS (best_tac (!claset addIs [r_into_rtrancl, rtrancl_trans])));
qed "Ap_reduce1";

goal Comb.thy "!!x z. x ---> y ==> z#x ---> z#y";
by (etac rtrancl_induct 1);
by (ALLGOALS (best_tac (!claset addIs [r_into_rtrancl, rtrancl_trans])));
qed "Ap_reduce2";

(** Counterexample to the diamond property for -1-> **)

goal Comb.thy "K#I#(I#I) -1-> I";
by (rtac contract.K 1);
qed "KIII_contract1";

goalw Comb.thy [I_def] "K#I#(I#I) -1-> K#I#((K#I)#(K#I))";
by (Fast_tac 1);
qed "KIII_contract2";

goal Comb.thy "K#I#((K#I)#(K#I)) -1-> I";
by (Fast_tac 1);
qed "KIII_contract3";

goalw Comb.thy [diamond_def] "~ diamond(contract)";
by (fast_tac (!claset addIs [KIII_contract1,KIII_contract2,KIII_contract3]) 1);
qed "not_diamond_contract";

(*** Results about Parallel Contraction ***)

(*Derive a case for each combinator constructor*)
val K_parcontractE = parcontract.mk_cases comb.simps "K =1=> z";
val S_parcontractE = parcontract.mk_cases comb.simps "S =1=> z";
val Ap_parcontractE = parcontract.mk_cases comb.simps "x#y =1=> z";

AddIs   parcontract.intrs;
AddSEs [K_parcontractE, S_parcontractE,Ap_parcontractE];
```

```
Addss  (!simpset);

(*** Basic properties of parallel contraction ***)

goal Comb.thy "!!x z. K#x =1=> z ==> (EX x'. z = K#x' & x =1=> x')";
by (Fast_tac 1);
qed "K1_parcontractD";
AddSDs [K1_parcontractD];

goal Comb.thy "!!x z. S#x =1=> z ==> (EX x'. z = S#x' & x =1=> x')";
by (Fast_tac 1);
qed "S1_parcontractD";
AddSDs [S1_parcontractD];

goal Comb.thy
 "!!x y z. S#x#y =1=> z ==> (EX x' y'. z = S#x'#y' & x =1=> x' & y =1=> y')";
by (Fast_tac 1);
qed "S2_parcontractD";
AddSDs [S2_parcontractD];

(*The rules above are not essential but make proofs much faster*)

(*Church-Rosser property for parallel contraction*)
goalw Comb.thy [diamond_def] "diamond parcontract";
by (rtac (impI RS allI RS allI) 1);
by (etac parcontract.induct 1 THEN prune_params_tac);
by (ALLGOALS Fast_tac);
qed "diamond_parcontract";

(*** Equivalence of x--->y and x===>y ***)

goal Comb.thy "contract <= parcontract";
by (rtac subsetI 1);
by (split_all_tac 1);
by (etac contract.induct 1);
by (ALLGOALS Fast_tac);
qed "contract_subset_parcontract";

(*Reductions: simply throw together reflexivity, transitivity and
  the one-step reductions*)

AddSIs [contract.K, contract.S];
AddIs [Ap_reduce1, Ap_reduce2, r_into_rtrancl, rtrancl_trans];

(*Example only: not used*)
```

```
goalw Comb.thy [I_def] "I#x ---> x";
by (Deepen_tac 0 1);
qed "reduce_I";

goal Comb.thy "parcontract <= contract^*";
by (rtac subsetI 1);
by (split_all_tac 1);
by (etac parcontract.induct 1 THEN prune_params_tac);
by (ALLGOALS (Deepen_tac 0));
qed "parcontract_subset_reduce";

goal Comb.thy "contract^* = parcontract^*";
by (REPEAT
    (resolve_tac [equalityI,
                  contract_subset_parcontract RS rtrancl_mono,
                  parcontract_subset_reduce RS rtrancl_subset_rtrancl] 1));
qed "reduce_eq_parreduce";

goal Comb.thy "diamond(contract^*)";
by (simp_tac (!simpset addsimps [reduce_eq_parreduce, diamond_rtrancl,
                                 diamond_parcontract]) 1);
qed "diamond_reduce";

writeln"Reached end of file.";
```

# Index